\documentclass[preprint,showpacs,preprintnumbers,amsmath,amssymb,nofootinbib,showkeys,aps]{revtex4}
\pdfoutput=1
\usepackage{graphicx}% Include figure files
\usepackage{dcolumn}% Align table columns on decimal point
\usepackage{bm}
\usepackage{epsfig}

\newcommand{\beq}{\begin{equation}}
\newcommand{\eeq}{\vspace{0cm} \end{equation}}
\newcommand{\beqq}{\setlength\arraycolsep{2pt}\begin{eqnarray}}
\newcommand{\eeqq}{\vspace{0cm} \end{eqnarray}}

\usepackage{hyperref}

\begin{document}
\title{New accelerating cosmology without dark energy:\\
The particle creation approach and the reduced relativistic gas}
%\title{ The particle creation approach to the reduced relativistic gas: New accelerating cosmology without dark energy}
\author{P. W. R. Lima$^{1}$\footnote{pablolima@usp.br}}
\author{J. A. S. Lima$^{1}$\footnote{jas.lima@iag.usp.br}}
\author{J. F. Jesus$^{2,3}$\footnote{jf.jesus@unesp.br}}
\affiliation{$^{1}$Departamento de Astronomia, Universidade de S\~{a}o
Paulo \\ Rua do Mat\~ao, 1226 - 05508-900, S\~ao Paulo, SP, Brazil}

\affiliation{$^2$Departamento de Ci\^encias e Tecnologia, Instituto de Ci\^encias e Engenharia, Universidade Estadual Paulista (UNESP) - R. Geraldo Alckmin 519, 18409-010, Itapeva, SP, Brazil
\\
$^3$Departamento de F\'isica, Faculdade de Engenharia e Ci\^encias de Guaratinguet\'a, Universidade Estadual Paulista (UNESP) - Av. Dr. Ariberto Pereira da Cunha 333, 12516-410, Guaratinguet\'a, SP, Brazil
}


\bigskip
\begin{abstract}
 The standard procedure to explain the accelerated expansion of the Universe is to assume the existence of an exotic component with negative pressure, generically called dark energy. Here, we propose a new accelerating flat cosmology without dark energy, driven by the negative creation pressure of a reduced relativistic gas (RRG). When the hybrid dark matter of the RRG is identified with cold dark matter, it describes the so-called CCDM cosmology whose dynamics is equivalent to the standard $\Lambda$CDM model at both the background and perturbative levels (linear and nonlinear). This effect is quantified by the creation parameter $\alpha$. However, when the pressure from the RRG slightly changes the dynamics of the universe, as measured by a parameter $b$, the model departs slightly from the standard $\Lambda$CDM cosmology.  Therefore, this two-parametric model ($\alpha, b$) describes a new scenario whose dynamics is different but close to the late-time scenarios predicted by CCDM and $\Lambda$CDM models. The free parameters of the RRG model with creation are constrained based on SNe Ia data (Pantheon+SH0ES) and also using $H(z)$ from cosmic clocks. In principle, this mild distinction in comparison with both CCDM or $\Lambda$CDM may help alleviate some cosmological problems plaguing the current standard  cosmology.
 
\end{abstract}

\maketitle

\section{Introduction}
It is widely believed that present day cosmology is dominated by dark energy (DE), an exotic fluid possessing negative pressure, introduced to explain the current accelerating stage of the universe, and also to solve the conflict between the low value of the mass density parameter (Cold Dark Matter + Baryons) and the high degree of flatness ($\Omega_{total}\sim 1$), as predicted by the inflationary paradigm.  The resulting model is supported by a plethora of observational results, including, among others, high redshift supernovae Ia \cite{supernovae, Suzuki:2011hu,Pantheon+}, estimates of the total age of the Universe \cite{AL199,CS2004}, galaxy clusters evolution \cite{Cluster1,Cluster2}, baryon acoustic oscillations \cite{SDSS2005,DESI2024}, cosmic background radiation \cite{Planck2019,Planck2020}, and cosmic chronometers \cite{JimenezLoeb02,MorescoEtAl20,MorescoEtAl22}.

Later, dozens of different dark energy candidates have been proposed in the literature \cite{review}. However, until now, the simplest and most theoretically compelling DE candidate is the cosmological constant ($\Lambda$) or, equivalently,  the vacuum energy density ($\rho_v = \Lambda/8\pi G$). At the level of Einstein Field Equations (EFE), it behaves like a rigid  substance (constant density) and negative pressure, obeying the equation of state (EoS), $p_v = -\rho_v$. The emerging cosmology is driven at late times by three conserved and noninteracting dominant components ($\Omega_B\sim 5\%, \Omega_{CDM} \sim 25\%, \Omega_{DE} \sim 70\%$), thereby defining the standard $\Lambda$CDM cosmology, also named cosmic concordance model. In addition, when combined with the very early inflationary scenario, it is usually considered to be pretty simple (spatially flat geometry), realistic, appealing, and predictive, provided that it is fine-tuned to fit the data. 

Although explaining most of the current astronomical observations, there are some old unsolved  theoretical ``puzzles" and also a couple of recent observational difficulties plaguing the $\Lambda$CDM model. In theoretical grounds, one may quote: (i) The old cosmological constant problem (CCP), and (ii) the cosmic coincidence\, ``mystery" (CCM). Basically, the $\Lambda$-problem is related with the extreme smallness of its present observed value. Actually, whether dark energy is represented by the vacuum state, the CCP constitutes one of the greatest challenges for our current understanding of fundamental Physics \cite{SW1989}. The CCM arises because the energy density of the rigid vacuum constant was astonishingly small in the radiation phase, but now it is finely tuned with the variable decreasing density of the dark matter component, or equivalently, $\Omega_\Lambda \sim \Omega_M$ at present \cite{PC03}. Furthermore, at least two observational difficulties has been recently added: (iii) the Hubble tension, and (iv) the $S_8$ tension, thereby showing persistent discrepancies of the $\Lambda$CDM predictions with the astronomical observations. 

The $H_0$ tension is an inconsistency between measurements of the Hubble constant ($H_0$), using Cepheid and calibrated supernovae of the SH0ES collaboration \cite{Riess2019} + strong lensing time delays \cite{B2019} and also the value $H_0 = 74$ km/s/Mpc from data at low and intermediate redshifts \cite{LV07}, while the Planck CMB power spectrum at high redshifts prefer values close to 68 km/s/Mpc. These results imply a statistical discrepancy around 5.4$\sigma$. The ongoing tension $S_8$ on the plane ($\Omega_M,\sigma_8$), where $\sigma_8$ is the current mass fluctuation on a scale of 8Mpc, arises when confronting  $\Lambda$CDM + Planck estimates with cosmic shear experiments \cite{kids}. 
%Currently, $H_0$ and $S_8$ tensions are challenging the $\Lambda$CDM model.

In this way, although $\Lambda$CDM being an interesting cosmic setting,  until now it seems like an incomplete description of the Universe. In certain sense, the cosmic concordance model is a comprehensible collection of ingredients brought together in order to explain the complete cosmic evolution of the Universe over all time and length scales. Such disparate ingredients are needed in order to smoothly connect the very early and late-time accelerating regimes \cite{DS2017}. Since the true nature of DE remains elusive, and, even the cosmic concordance model is not free of problems, new possibilities are being carefully scrutinized, mainly the ones slightly departing from $\Lambda$CDM cosmology. Some examples are the research line started long ago by decaying vacuum models \cite{LH}, just the first kind of models assuming interactions in the dark sector \cite{Inter}, cosmologies based on modifications of Einstein’s gravity \cite{SF2010}, and also models reducing the dark sector and whose dominant component is  dark matter (see below). 

Actually, part of the $\Lambda$CDM problems inspired some authors to explore the possibility of reducing the dark sector by taking $\Omega_{DE}\equiv 0$. The basic idea is that different from DE, the DM component is needed in practically all relevant scales (from galaxies to cosmology). Such an approach quite investigated recently, is based on the assumption that the accelerating fuel is related to a negative pressure provided by the matter creation process in the evolving Universe \cite{LSS2008,LBC2012}. This reduction in the dark sector is quite interesting because it replaces the idea of an unknown component by a cosmic mechanism of microscopic origin (matter-creation), and as such, bringing new possibilities to solve the problems plaguing the standard model. In particular, it has been shown that the so-called creation cold dark matter cosmology (CCDM), sometimes referred to as LJO model, emulates the $\Lambda$CDM cosmology with just one free parameter $\alpha$ defining the creation rate \cite{LJO2010}. Interestingly, the equivalence is not restricted to the background solution, since it also happens at the perturbative linear and non-linear levels \cite{Pert1,Pert2}. Even so, apart the CCP and CCM puzzles which are absent in such scenario, the CCDM model may share the same observational problems of the $\Lambda$CDM model. However, is has been argued that such model (or some variant of it) may solve in a natural way, the theoretical and observational problems of $\Lambda$CDM \cite{TL2023,EKO2024}.
%In addition to the dark energy problem, the traditional $\Lambda$CDM scenario suggests that approximately 26\%  of the universe is composed of another component in the so-called dark sector, a fluid dubbed cold dark matter (CDM), whose nature is also  unknown by the standard model of particle physics. 

In this connection, we recall that some attention has also been paid to a new warm dark matter (WDM) component, the so-called reduced relativistic gas (RRG). Its possible relevance to cosmology has also been investigated. In particular,   %\cite{Peixoto2005,Fabris2009,Fabris2012,Medeiros2012}. 
it has been shown that RRG models provide relativistic corrections to the dynamics of the universe, and this kind of warm dark matter (WDM) particles allows large-scale structure (LSS) to be formed as we observe today \cite{Hipolito2018,Pordeus2019,Pordeus2021}. 
Nevertheless, since the RRG at late time behaves exactly like a cold dark-matter component, the late time accelerating stage cannot be driven by the RRG model. So, as far as we know, a dark energy component has usually been adopted in the literature of RRG cosmology (see also Section III). 
%\cite{Pordeus2019,Pordeus2021,Hipolito2018}. 
%li\cite{Peixoto2005,Fabris2009,Fabris2012,Medeiros2012
%Pordeus2019,Pordeus2021,Hipolito2018}.

In this paper, we propose a model in which the fuel accelerating the universe is generated by the gravitationally induced particle creation of the RRG component. More precisely, it is assumed that WDM particles spring up into  spacetime  at the same rate as proposed in the CCDM scenario, while the EoS of this gas is described within the so-called RRG approximation.  This approach not only describes an accelerated expansion phase at late times of the universe without dark energy, but also accounts for relativistic contributions from the RRG gas particles in the early  dynamics of the universe. In this way, the CCDM model which mimicks the $\Lambda$CDM models is slightly modified. These relativistic contributions can be quantified by the warmness parameter $b$, which is the ratio between the relativistic and non-relativistic energy densities of the massive particles. In order to constrain the values of the pair of free parameters  ($\alpha,b$) we analyse apparent magnitudes from type Ia supernovae (SNe Ia) combined with Cepheid distances from the Pantheon+SH0ES data set and $H(z)$ measurements from cosmological clocks. 

The paper is organized as follows. In Section II,  the basics of macroscopic particle production in cosmology and the associated dynamics are briefly discussed.  Section III introduces the equation of state approximation proposed long ago for the reduced relativistic gas (RRG). Section IV discusses the model in which WDM particles are gravitationally created at the production rate proposed by the CCDM model. In Section V, we present our statistical analyses and constraints to the free parameters of the model. The main findings of the paper are summarized in Section VI. Finally, in Appendix A, all details describing the transition redshift are presented. 

\section{Cosmic Dynamics and Thermodynamics with Particle Production}

In this section, we briefly review the cosmic dynamics and thermodynamics for a single fluid endowed with `adiabatic' particle production, focusing especially on those aspects relevant to the RRG component that will be discussed next section. 

To begin with, let us consider the space–time described by a
flat Friedmann-Lema\^itre-Robertson-Walker (FLRW) geometry:
\begin{equation}
ds^{2} = dt^{2} - a^{2}(t) \left[dx^{2} + dy^{2} + dz^{2}\right],
\label{ds2}
\end{equation}
where $a(t)$ is the scale factor.

For now we will ignore the baryonic contribution. So, the cosmic dynamics is driven by the RRG gas endowed with the macroscopic description of particle production put forward by Prigogine \textit{et al.} \cite{PRI89} and extended through a manifestly covariant approach by Calv\~ao \textit{et al.} \cite{CLW92}. Later, it was shown that the matter creation process is completely different from the bulk viscosity \cite{LG92}, an effective mechanism proposed long ago by Zeldovich \cite{zeld70} and also adopted by several authors to describe the matter creation process \cite{PZ93}. 

The matter-creation mechanism is assumed here to be the unique source of acceleration acting in the current stage of the universe.  Its creation pressure ($p_c$)  provides the macroscopic phenomenological description of the quantum back reaction effect on the geometry associated with the gravitational creation process induced by the expanding universe. This basic creation effect has a quantum origin, as many authors have discussed in the literature \cite{PT}.   

In this case, the Einstein Field Equations (\textbf{EFE}) for a single fluid endowed with matter-creation can be written as (for a multifluid component with matter creation see \cite{Pert2,TL2023})
\begin{align}
8\pi G \rho &= 3H^2,\label{EE1}\\
8\pi G (p + p_c) &= -2\dot{H} - 3H^{2},\label{EE2}
\end{align}
where an overdot means time comoving derivative, $H=\dot{a}/a$ is the Hubble parameter, $\rho$, $p$ and $p_c$ are the energy density, the equilibrium pressure, and creation pressure, respectively. 

The thermodynamic states of a relativistic fluid are defined by the energy conservation law (ECL), which is also contained in the \textbf{EFE} ($u_{\mu} T^{\mu\nu}_{;\nu}= 0$), and the conservation of the particle and entropy fluxes $N^{\mu}_{;\mu}=0$ and $S^{\mu}_{;\mu}=0$. However, in the presence of gravitationally induced particle creation, the ECL contained in the \textbf{EFE}, must be complemented by the balance equations for $N^{\mu}$ and $S^{\mu}$, respectively. In this case, one may write: 

\begin{eqnarray}\label{TEQ}
u_{\mu} T^{\mu\nu}_{;\nu}&=& 0 \,\,\,\, \Longleftrightarrow \,\,\,\, \dot{\rho} + 3H(\rho + p + p_c)=0,   \label{n_eq} 
\\
N^{\mu}_{;\mu} &=& n\Gamma \,\,\,\, \Longleftrightarrow \,\,\,\,\ \dot n + 3nH = n\Gamma
 \Longleftrightarrow \frac{\dot N}{N} = \Gamma,  \label{n_gamma}
\\ 
S^{\mu}_{;\mu} &=& s\Gamma \,\,\,\, \Longleftrightarrow \,\,\,\,\ \dot s + 3sH = s\Gamma\,\, \Longleftrightarrow \frac{\dot S}{S} = \Gamma,   \label{s_gamma}
\end{eqnarray}
where $n$ and $s$ are the particle concentration and entropy density and $\Gamma$ is the particle creation rate.  The quantities $N=na^3$ and $S=sa^{3}$, are the total number of particles and entropy in a comoving volume, respectively. The concept of\, `adiabatic' creation is very simple to understand. From Eqs. \ref{n_gamma} and \ref{s_gamma} we see that

\begin{equation}
    \frac{\dot S}{S} = \frac{\dot N}{N}  = \Gamma \geq 0,
\end{equation}
where the inequality is a consequence of the second law of thermodynamics. The positiviness of $\Gamma$ means that the spacetime may only create matter (entropy). The natural solution is $S=k_B N$, where $k_B$ is the boltzmann constant.  In addition, since the entropy per particle is $\sigma = S/N$, the first equality above means that
\begin{equation}
   \dot \sigma =  \sigma \left(\frac{\dot S}{S}-\frac{\dot N}{N}\right) = 0.  
\end{equation}
Hence,\, `adiabatic' particle creation means the total entropy $S$ and  number of particles $N$ increases, but  the specific entropy remains constant ($\dot\sigma=0$). 

It is easy to see that thermodynamics also determines the expression of the creation pressure. The thermodynamic quantities are related by the local Gibbs law: $nTd\sigma =  d\rho - (\rho + p)\frac{dn}{n}$, where T is the temperature. So, by using that $\dot\sigma=0$ and combining that result with (\ref{n_eq}) and (\ref{n_gamma}), it is readily seen that \cite{LJO2010,LBC2012,Pert1,Pert2,TL2023,EKO2024}
\begin{equation}\label{Pc1}
p_{c}=-(\rho + p)\, \frac{\Gamma}{3H}\,.
\end{equation}
In the `adiabatic condition', $\Gamma$ is positive definite. Therefore, the creation pressure may be always negative provided that $\rho+p$ is positive. This happens, for instance,  by assuming a normal equation of state (EoS), that is, without any kind dark energy
\begin{equation}\label{EoS}
p=\omega\rho, \,\,\,\omega \geq 0.
\end{equation}

Now, by combining Eq. (\ref{n_eq}) with (\ref{Pc1}) and (\ref{EoS}), one obtains the differential equation governing the evolution of the energy density:
\begin{equation}\label{ECO2}
\dot{\rho}+3H\rho(1+\omega)\left(1-\frac{\Gamma}{3H}\right)=0 \, .
\end{equation}

%
%  it is readily checked from (1)-(4) that  the evolution of the scale factor is governed by the differential equation 
% \begin{equation}\label{GE1}
% a\ddot{a}+\left(\frac{1+3\tilde\omega}{2}\right)\dot{a}^{2}+\left(\frac{1+3\tilde\omega}{2}\right)k=0 
% \end{equation}
% where $\tilde\omega$ is the effective value of the EoS parameter
% modified by the particle creation pressure
% \begin{equation}\label{EEoS}
% \tilde\omega_\equiv \frac{P_{total}}{\rho} = \frac{p+p_c}{\rho}=\omega -\left(1 +\omega\right)\frac{\Gamma}{3H}.
% \end{equation}

In the absence of particle creation ($\Gamma = 0$), equation (\ref{ECO2}) reduces to the flat FLRW equation. For $\Gamma = 3H$, the above equation yields $\dot{\rho}=0$, that is, $\rho$ constant,  regardless of the value of $\omega$. In other words,  `adiabatic' particle production allows de Sitter space-times whose matter-energy content is supported by a pressureless fluid or even radiation \cite{PRI89,CLW92}. Next section, we shall consider the possibility that the gas of created particles is a reduced relativistic gas (RRG).

%%%%%%%%%%%%%%%%%%%%%%%%%%%%%%%%%%%%%%%%%%%%%%%%%%%%%%%%%%%%%%%%%%%%%%%%%%%%%%%%%
%%%%%%%%%%%%%%%%%%%%%%%%%%%%%%%%%%%%%%%%%%%%%%%%%%%%%%%%%%%%%%%%%%%%%%%%%%%%
\section{Reduced Relativistic Gas (RRG)}
%%%%%%%%%%%%%%%%%%%%%%%%%%%%%%%%%%%%%%%%%%%%%%%%%%%%%%%%%%%%%%%%%%%%%%%%%%%%%%%%
%%%%%%%%%%%%%%%%%%%%%%%%%%%%%%%%%%%%%%%%%%%%%%%%%%%%%%%%%%%%%%%%%%%%%%%%%%%%%%%%%%%%
 
Contributions from WDM particles to cosmology, along with other implications to astronomy, have increased the interest in the relativistic behavior of dark matter particles \cite{PDU1,PDU2,Horiuchi2016}. 
The appropriate  EoS for an ideal  gas of relativistic massive particles was derived long ago \cite{Jutner1911} 
\begin{equation}\label{bessel}
    \rho = \frac{K_{3}(\rho_{m} /P)}{K_{2}(\rho_{m} /P)}\rho_{m} - P,
\end{equation}
where $\rho_{m}=nmc^2$ is the rest energy density of a gas with particle number density $n=N/V$, while $K_{2}$ and $K_{3}$ are modified Bessel functions of the second kind. 

The EoS (\ref{bessel}) leads to a complicated differential equation governing the evolution of the cosmos. On the other hand, an alternative description of a relativistic gas was proposed by Sakharov in which a phenomenological equation of state describes a fluid that transits from an ultra-relativistic to a pressureless matter behavior as the universe expands \cite{sakharov1966}. Later, an approximation to the relativistic description of the ideal gas was obtained explicitly under the hypothesis that identical particles of the gas share equal momentum magnitudes \cite{Peixoto2005}. 

In this new approach, the EoS describing the so-called Reduced Relativistic Gas (RRG) takes the following form:  
\begin{equation}\label{EoSRRG}
    P=\frac{\rho}{3}\left[1-\frac{\rho_{mo}^{2}}{\rho^{2}}\left(\frac{n}{n_o}\right)^2\right],
\end{equation}
where $p$ and $\rho$ are, respectively, the pressure and energy density of the gas and $\rho_{mo}=n_om$ is the rest energy density of massive particles ($\rho_{m}=nm$ evaluated at the present time ($t=t_o$). 

As it appears, the above EoS is much simpler than equation (\ref{bessel}) and allows analytical solutions to be obtained for models in which the relativistic contributions of massive particles are significant in the dynamics of the universe. Moreover, it was shown \cite{Peixoto2005,Fabris2009} that the RRG equation of state is a very efficient approximation to the EoS (\ref{bessel}). Note also that it depends explicitly on $n$ so that a third differential equation is required to solve the Einstein Equations (\ref{EE1}) and (\ref{EE2}). Accordingly, the evolution of the particle concentration as given by the balance equation (\ref{n_gamma}) must be used.
%The particle concentration changes the dynamics of the universe whether the EoS of a fluid depends explicitly %on $n$, i.e. $\rho=\rho(P,n)$, or the number of particles varies in time ($\dot{N}\neq 0$) (Citar Ademir). 
%%%%%%%%%%%%%%%%%%%%%%%%%%%%%%%%%%%%%%%%%%%%%%%%%%%%%%%%%%%%%%%%%%%%%%%%%%%
%\begin{equation}
%\frac{\dot{n}}{n}+3H=\Gamma.
%\label{ndot}
%\end{equation}
Conversely,  when the number of particles in a comoving volume ($N=na^{3}$) is conserved,  $n \propto a^{-3}$ and $\Gamma=0$ so that from (\ref{Pc1}) the creation pressure is also nullified ($p_c=0$). Thus, the cosmic equations (\ref{EE1}) and (\ref{EE2}) can be easily solved with the help of EoS (\ref{EoSRRG}). In this case EoS (\ref{EoSRRG}) can also be used to rewrite  (\ref{ECO2}) as: 
%%%%%%%%%%%%%%%%%%%%%%%%%%%%%%%%%%%%%%%%%%%%%%%%%%%%%%%%%%%%%%%%%%%%%%%%%%%%%%%%%%%%%%%%%%%%%%%%%%%%
\begin{equation} \label{rmevol}
\frac{d\rho}{da}=-3\frac{\rho}{a}\left[\frac{4}{3}-\frac{1}{3}\frac{\rho_{mo}^{2}}{\rho^{2}} \left(\frac{a_o}{a}\right)^6  \right],
\end{equation}
%%%%%%%%%%%%%%%%%%%%%%%%%%%%%%%%%%%%%%%%%%%%%%%%%%%%%%%%%%%%%%%%%%%%%%%%%%%%%%%%%%%%%%%%%%%%%%%%%%%%%
 whose solution is given by
%%%%%%%%%%%%%%%%%%%%%%%%%%%%%%%%%%%%%%%%%%%%%%%%%%%%%%%%%%%%%%%%%%%%%%%%%%%%%%%%%%%%%%%%%%%%%%%%%%%%%
\begin{equation}\label{rmRRG}
\rho(a) = \sqrt{\rho_{mo}^{2}\left(\frac{a_{o}}{a}\right)^{6}+\rho_{ro}^{2}\left(\frac{a_{o}}{a}\right)^{8}},
\end{equation}
%%%%%%%%%%%%%%%%%%%%%%%%%%%%%%%%%%%%%%%%%%%%%%%%%%%%%%%%%%%%%%%%%%%%%%%%%%%%%%%%%%%%%%%%%%%%%%%%%%%%%%
%$\rho(a_o)=\sqrt{\rho_{no}^{2} + \rho_{ro}^2}$ is the energy density (\ref{rmRRG}) evaluated today ($a=a_o$) and
where $\rho_{ro}$ is an integration constant whose interpretation is a simple one: when $a \ll a_o$, the second term dominates over the first and the energy density behaves as $\rho\propto a^{-4}$, which has the same evolution as radiation. Eventually, as $a(t)$ increases, the second term becomes negligible compared to the first and $\rho$ becomes $\rho \propto a^{-3}$, which is consistent with the behavior of pressureless matter at late times. It is also convenient to write $\rho$ as 

\begin{equation}\label{rmRRGb}
\rho = \rho_{mo}\left(\frac{a_{0}}{a}\right)^{3} \sqrt{1+b^{2}\left(\frac{a_{0}}{a}\right)^{2}},
\end{equation}
where $b=\rho_{ro}/\rho_{mo}$ is known as the warmness parameter. This parameter is important in determining the relevance of relativistic contributions to energy density as the universe expands. 
%For instance, if $b\ll1$ the relativistic behaviour of the gas can be negligible at $a \sim a_o$ yet significant at $a\ll a_o$. 

% Moreover, it was shown that if the gas was in thermal equilibrium with radiation in the early universe, than the mass of the particles are related to $b$ by $m\sim 10^{-7}b^{-1}$ keV \cite{pordeus2019}. However, the mass of dark matter particles, which is the most abundant type of massive particles in the universe, is still unknown. Recent research indicates that $m_{dm}\sim 10^{-3} - 10^{7}$, which leads to a range $b\sim 10^{-14} - 10^{-4}$ \cite{calmet2021}. 

Several models have considered relativistic contributions from the RRG approximation to cosmology. Solutions for models with pure WDM, WDM with radiation, WDM + $\Lambda$ and others multifluids have been obtained in the literature both in background and perturbative levels of cosmology assuming the EoS (\ref{EoSRRG}) 
%\cite{Peixoto2005, Fabris2009, , 
\cite{Fabris2012, Pordeus2019, Pordeus2021,Slepian2018}. Other approaches such as running vacuum cosmologies, alternative space-time geometry and scalar fields have also been studied in the context of RRG \cite{Reis2018,Ruiz2020}. In the next Section we will investigate a scenario where the particles of the RRG gas are allowed to be created from the mechanism of gravitationally induced particle creation. As remarked in the Introduction, the derived  accelerating cosmology has no dark energy $(\Omega_{DE}=0)$.

%%%%%%%%%%%%%%%%%%%%%%%%%%%%%%%%%%%%%%%%%%%%%%%%%%%%%%%%%%%%%%%%%%%%%%%%%%%%%%%%%%%%%%%%%
%%%%%%%%%%%%%%%%%%%%%%%%%%%%%%%%%%%%%%%%%%%%%%%%%%%%%%%%%%%%%%%%%%%%%%%%%%%%%%%%%%%%%%%%%%%%
\section{Modified CCDM: RRG Model with Particle Creation}\label{RRG_CCDM}  
%%%%%%%%%%%%%%%%%%%%%%%%%%%%%%%%%%%%%%%%%%%%%%%%%%%%%%%%%%%%%%%%%%%%%%%%%%%%%%%%%%%%%%%%%%%%%%
%%%%%%%%%%%%%%%%%%%%%%%%%%%%%%%%%%%%%%%%%%%%%%%%%%%%%%%%%%%%%%%%%%%%%%%%%%%%%%%%%%%%%%%%%%%%%%%
Let us now consider a model endowed with the ``adiabatic'' mechanism of warm particle creation. As discussed in the Introduction, the particle production rate $\Gamma$ is fundamental in determining the cosmological parameters in models that $\dot{N}\neq0$. Rigorously, $\Gamma$ should be determined from quantum field theory (QFT) of non-equilibrium in curved spacetimes. However, this theory has yet to be formulated. On the other hand, the particle production rate can be obtained phenomenologically. Similarly to the creation rate proposed in \cite{LJO2010}, we investigate the case where $\Gamma$ is given by 

%%%%%%%%%%%%%%%%%%%%%%%%%%%%%%%%%%%%%%%%%%%%%%%%%%%%%%%%%%%%%%%%%%%%%%%%%%%
\begin{equation}\label{gammadm}
\Gamma=3\alpha \left( \frac{\rho_{o}}{\rho_{m}} \right) H,
\end{equation}
%%%%%%%%%%%%%%%%%%%%%%%%%%%%%%%%%%%%%%%%%%%%%%%%%%%%%%%%%%%%%%%%%%%%%%%%%%%
where $\alpha$ is a dimensionless constant, $\rho_{o}$ is the critical energy density evaluated today, 3 is only a mathematical convenience and $\rho_{m}$ is the rest energy density $\rho_{m} = nmc^2$ of dark matter and baryon particles $\rho_m=\rho_{dm}+\rho_{b}$. As one may recall from the previous section, the particle concentration $n$ can be solved in terms of the scale factor from the balance equation (\ref{n_eq}) and the production rate (\ref{gammadm}) yields:
%%%%%%%%%%%%%%%%%%%%%%%%%%%%%%%%%%%%%%%%%%%%%%%%%%%%%%%%%%%%%%%%%%%%%%%%%%%%%%%%%%%%%%%%%%%%%%
\begin{equation} \label{complete}
\frac{n}{n_o}=\frac{\rho_{m}}{\rho_{mo}}= \left( 1 - \alpha \frac{\rho_{o}}{\rho_{mo}}  \right) \left(\frac{a_o}{a}\right)^{3} + \alpha \frac{\rho_{o}}{\rho_{mo}}.
\end{equation}
It is clear from this equation that $\rho_m$ can be written as
\begin{equation}\label{rmLJO}
\rho_{m} = (\rho_{mo} - \alpha \rho_{o})\left(1+z\right)^{3}+ \alpha \rho_{o},
\end{equation}
which is exactly the solution for the matter energy density of the CCDM model \cite{LJO2010}.

If the particle creation rate is neglected, that is,  $\alpha=0$, the standard case $n\propto\rho_{m}\propto a^{-3}$ is recovered from the above equation, as expected. From the particle production rate (\ref{gammadm}) and the particle concentration (\ref{complete}) one can write the ECL for the RRG gas modified by the mechanism of particle creation
%The conservation of energy can then be written as 
%%%%%%%%%%%%%%%%%%%%%%%%%%%%%%%%%%%%%%%%%%%%%%%%%%%%%%%%%%%%%%%%%%%%%%%%%%%%%%%%%%%
%\begin{equation}
%\dot{\rho}=\left(\rho+p\right)\frac{\dot{n}}{n}
%\end{equation}
%%%%%%%%%%%%%%%%%%%%%%%%%%%%%%%%%%%%%%%%%%%%%%%%%%%%%%%%%%%%%%%%%%%%%%%%%%%%%%%%%%%%
%%%%%%%%%%%%%%%%%%%%%%%%%%%%%%%%%%%%%%%%%%%%%%%%%%%%%%%%%%%%%%%%%%%%%%%%%%%%%%%%%%%%%%%%%%%%%%%%%%%%
\begin{equation} \label{rmevol2}
\frac{d\rho}{da}=-3\frac{\rho}{a}\left( 1 - \alpha \frac{\rho_{o}}{\rho_{m}}\right)\left[\frac{4}{3}-\frac{1}{3}\frac{\rho_{m}^{2}}{\rho^{2}}  \right],
\end{equation}
%%%%%%%%%%%%%%%%%%%%%%%%%%%%%%%%%%%%%%%%%%%%%%%%%%%%%%%%%%%%%%%%%%%%%%%%%%%%%%%%%%%%%%%%%%%%%%%%%%%%%
where $\rho$, in this case, is the energy density of the reduced relativistic gas of massive particles. Note that equation (\ref{rmevol}) is a particular case of the ECL (\ref{rmevol2}) when $\alpha=0$.

The differential equation in (\ref{rmevol2}) can be extremely simplified if the scale factor and its differential are written in terms of $\rho_{m}$ and other constants. Note that differentiating equation (\ref{rmLJO}) with respect to $a$ yields 
\begin{equation}\label{da}
   \frac{da}{a}= - \frac{1}{3} \left(\frac{a}{a_o}\right)^3 \frac{d\rho_m}{\rho_{mo} - \alpha \rho_{o}},  
\end{equation}
and, also from equation (\ref{rmLJO}), note that
\begin{equation}\label{a3}
     \left(\frac{a}{a_o}\right)^{3}=\frac{\rho_{mo} - \alpha \rho_{o}}{\rho_{m} - \alpha \rho_{o}}.
\end{equation}
Now, by  inserting equations (\ref{da}) and (\ref{a3}) into (\ref{rmevol2}) the ECL becomes simply 
\begin{equation}
    \frac{d\rho}{d\rho_{m}}= \frac{4}{3} \frac{\rho}{\rho_{m}} - \frac{1}{3} \frac{\rho_{m}}{\rho}.
\end{equation}
This first-order differential equation is easily solved. Introducing a new variable, $u\equiv\rho/\rho_m$, the solution reads: 
%It is also convenient to redefine the variables $x=\rho_{m}$, $y=\rho_{n}$ so that the equation above can be expressed as a homogeneous differential equation of order zero, so it reads
%\begin{equation}
%    \left(\frac{x}{3y}-\frac{4y}{3x} \right)dx + dy = 0.
%\end{equation}
\begin{equation}\label{u}
    u = \sqrt{ 1 + (C\rho_m)^{2/3}}, 
\end{equation}
where $C$ is a constant of integration. As $\rho=\rho_mu$, it follows that 
% Thus, substituting $\rho_{m}$ and $\rho_{n}$ back into $u$ and $x$, it reads
\begin{equation}
    \rho=\sqrt{\rho_{m}^2 + \left( \rho_{o}^2 - \rho_{mo}^2 \right) \left(\frac{\rho_{m}}{\rho_{mo}}\right)^{8/3}},
\end{equation}
in which $C=(\rho_{o}^2 - \rho_{mo}^2)/\rho_{m}^{8/3}$ can be obtained as an initial condition for $\rho$ at $a=a_o$ in (\ref{u}). Finally, by defining $\rho_{ro}=\sqrt{\rho_{o}^2 - \rho_{mo}^2}$ as the relativistic contribution of the massive particles to the energy density 
%%%%%%%%%%%%%%%%%%%%%%%%%%%%%%%%%%%%%%%%%%%%%%%%%%%%%%%%%%%%%%%%%%%%%%%%%%%%%%%%%%%%%%%%%%%%%%%%%%%%%
\begin{equation} 
\rho=\sqrt{\rho_{mo}^{2}\left(\frac{\rho_{m}}{\rho_{mo}}\right)^{2}+\rho_{ro}^{2}\left(\frac{\rho_{m}}{\rho_{mo}}\right)^{8/3}} = \rho_{m} \sqrt{1+b^{2}\left(\frac{\rho_{m}}{\rho_{mo}}\right)^{2/3}},
\end{equation}
%%%%%%%%%%%%%%%%%%%%%%%%%%%%%%%%%%%%%%%%%%%%%%%%%%%%%%%%%%%%%%%%%%%%%%%%%%%%%%%%%%%%%%%%%%%%%%%%%%%%%
where $b=\rho_{ro}/\rho_{mo}$ is the warmness parameter. In terms of the redshift, $\rho$ can be rewritten as
% where $\rho_{ro}$ is the constant of integration regarding the relativistic behavior of the dark matter particles, and $\rho_{mo}=\sqrt{\rho_{no}^{2} + \rho_{ro}^{2}}$ is the initial condition at $z=0$. 

\begin{equation}\label{rmz}
\rho=  \left[\left(\rho_{mo}- \alpha \rho_{o}\right)\left(1+z\right)^{3} + \alpha \rho_{o} \right]\sqrt{ 1 + b^{2}\left[\left(1- \frac{\alpha \rho_{o}}{\rho_{mo}}\right)\left(1+z\right)^{3} + \frac{\alpha \rho_{o}}{\rho_{mo}}  \right]^{8/3}}.
\end{equation}
Note that in the absence of particle production ($\alpha=0$), the equation above recovers the reduced relativistic gas from (\ref{rmRRG}). In addition,  for a gas with negligible warmness parameter ($b=0$), that is, a gas with non-relativistic particles, the solution (\ref{rmz}) reduces to Eq. \eqref{rmLJO}, as expected.
 
Let us now determine the contributions of the warm particle creation to the dynamics of the universe. Considering that the only component of the universe being the gas described above,
the energy density (\ref{rmz}) coincides with the critical energy density in (\ref{EE1}). Thus, the Hubble parameter for this model can be obtained by substituting equation (\ref{rmz}) into (\ref{EE1}) and, considering $\Omega_{m}=\rho_{mo}/\rho_{o}$, it reads  
\begin{equation}\label{Hz}
H^2= H^{2}_{o}\left[\left(\Omega_{m} - \alpha \right)\left(1+z\right)^{3} + \alpha\right]\sqrt{ 1 + \frac{b^2}{\Omega_{m}^{2/3}} \left[ \left(\Omega_{m} - \alpha\right)\left(1+z\right)^{3} + \alpha \right]^{2/3}}.
\end{equation}

In models which the acceleration of the universe is due to the mechanism of particle creation, the late phase of the universe is dominated by matter, $\Omega_m=1$. However, galaxy cluster observations and other probes suggest that $\Omega_m \approx 0.3$ \cite{Cluster1,Rapetti2008,Rod2019,Q2024}. The explanation for the low $\Omega_m$ constraints in models with particle creation is that the effective matter energy density, which scales as $(1+z)^3$ and is responsible for the structure formation, is $\Omega_{meff}=\Omega_{m}-\alpha$. In this point of view, the matter that did not agglomerate is homogeneously distributed across the universe \cite{LJO2010}. This is similar to the model proposed in this paper (see equation \eqref{Hz}), except it is modified by relativistic contributions.

Additionally, by evaluating $H(z)$ at $z=0$ in equation (\ref{Hz}), we see that $ \Omega_{m}= 1/\sqrt{1+b^2}$. Thus, the Hubble parameter can be written only in terms of $H_0$, $\alpha$ and $b$ as 
\begin{equation}\label{Hzs}
H^2=  H^{2}_{o} \left[\left(\frac{1}{\sqrt{1+b^{2}}}-\alpha\right)\left(1+z\right)^{3}+\alpha\right]\sqrt{1+b^{2}\left(1+b^{2}\right)^{1/3}\left[\left(\frac{1}{\sqrt{1+b^{2}}}-\alpha\right)\left(1+z\right)^{3}+\alpha\right]^{2/3}}.
\end{equation}
%where $\alpha_*=\alpha\sqrt{1+b^2}$.
% \begin{equation}\label{H(a)late}
% \frac{H^2}{H^{2}_{o}}=   \left[\left(\frac{1}{\sqrt{1+b^2}} - \alpha\right)\left(1+z\right)^{3} + \alpha\right]\sqrt{ 1 + b^2 \left(1+b^2\right)^{1/3} \left[ \left(\frac{1}{\sqrt{1+b^2}} - \alpha\right)\left(1+z\right)^{3} + \alpha \right]^{2/3}},
% \end{equation}
%where the only free parameters are $\alpha$, $\Omega_{dm}$ and $H_0$. Note that the effective matter energy fraction is $\Omega_{meff}=\Omega_{dm} - \alpha$. Also, 
%It is clear in the expression above that $b=0,\alpha\neq0$ one recovers the LJO model from \cite{LJO10}, and that for $\alpha=0,b\neq0$ the RRG model from \cite{peixoto2005} is recovered. 
The effective matter energy density is now a function of $\alpha$ and $b$, and can be written as $\Omega_{meff}=(1/\sqrt{1+b^2})-\alpha$. Note from the equation above that as $z\rightarrow -1$ ($a\rightarrow \infty$), the expansion rate $H_f=H(a\rightarrow \infty)$ becomes approximately constant  
\begin{equation}
H_{f}\approx H_o \sqrt{\alpha} \left[  1+ b^2 \left(1+b^{2}\right)^{1/3} \alpha^{2/3}  \right]^{1/4},    
\end{equation}
 which implies an exponential expansion in the very late universe since $\dot{a}/a=H_{f}$ yields ($a\propto e^{H_{f} t}$). This dynamic behavior at $a\rightarrow \infty$ is consistent with both the CCDM ($H_{f}=H_o \sqrt{\alpha}$) and $\Lambda$CDM ($H_{f}=H_o \sqrt{\Omega_{\Lambda}}$) models, except for different values of $H_{f}$. 
 
Several authors have discussed the degeneracy of the dynamics of the universe between the $\Lambda$CDM model and models with non-relativistic particle creation, for $\Gamma$ in the form (\ref{gammadm}), if $\alpha=\Omega_{\Lambda}$. On the other hand, the dynamics of the universe in a model with RRG is clearly different from the one in the $\Lambda$CDM model. Nonetheless, these relativistic corrections might solve or alleviate some problems of the standard model. In the next section, we confront the model against late-time cosmological observational data in order to constrain its free parameters.
%Thus, one can say that the model described above is ``quasi''-$\Lambda$CDM.

\section{Analysis and Results}\label{analysis}

In order to constrain the free parameters of the RRG model with particle creation, we make use of the following observational data: apparent magnitudes from SNe Ia combined with Cepheid distances, as given by the Pantheon+SH0ES compilation \cite{Pantheon+} and $H(z)$ from cosmic chronometers \cite{MorescoEtAl20,MorescoEtAl22}.

\subsection{Data and Methodology}
The Pantheon+SH0ES compilation \cite{Pantheon+} comprises currently one of the largest SNe Ia compilations, combining Pantheon+ SNe Ia with Cepheid distances from SH0ES. The Pantheon+ consists of 1701 light curves of 1550 distinct SNe Ia, spanning a redshift interval $0.001<z<2.26$. By using this compilation, it is possible to constrain the free parameters of the model including $H_0$, which is obtained from the SH0ES calibration.

Besides being a large compilation yielding powerful constraints, the Pantheon+SH0ES also is independent of the cosmological model choice, depending only on the SNe Ia and Cepheid Astrophysics, being then suitable for constraining the RRG cosmological model.

Another dataset which is independent of the cosmological model choice and dependent only on astrophysical assumptions, is the $H(z)$ data from cosmic chronometers (CC). $H(z)$ data obtained this way depends only on chemical evolution models, from which ages can be obtained, so $H(z)$ can be obtained through the relation $H(z)=-\frac{1}{(1+z)dt/dz}$ \cite{JimenezLoeb02}.

We use the 32 CC $H(z)$ data compiled from Moresco et al. \cite{MorescoEtAl20,MorescoEtAl22}. This compilation includes the first treatment of the CC systematic errors, which increased the errors when compared with previous compilations, but yields more robust constraints.

Following the Bayes' Theorem, in both datasets, we aim to probe the posterior $p$ of the parameters $\theta=(M,H_0,\alpha,b)$ given the data, $D$:
\begin{equation}
    p(\theta|D)\propto\pi(\theta)\mathcal{L}(\theta|D),
\end{equation}
where $\pi$ is the prior, which we choose to be flat for all parameters and $\mathcal{L}$ is the likelihood, given by
\begin{equation}
    \mathcal{L}=\mathcal{L}_{SN}\mathcal{L}_H,
\end{equation}
where $\mathcal{L}_{SN}$ is the likelihood of the Pantheon+SH0ES compilation and $\mathcal{L}_{H}$ is the CC likelihood. Both likelihoods can be written as:
\begin{equation}
    \mathcal{L}_D\propto e^{-\chi_D^2/2},
\end{equation}
where $D=(SN,H)$ and
\begin{equation}
    \chi_D^2=\sum_{i,j}(y(z_i,\theta)-y_{obs,i})C_{ij}^{-1}(y(z_j,\theta)-y_{obs,j}),
\end{equation}
where $C_{ij}$ is the covariance matrix and the data $y_i$ is the distance modulus $\mu_i$ for SNe Ia and $H_i$ for CC $H(z)$ data.

In order to probe the posteriors, we used the python package emcee \cite{ForemanMackey13}, which generates Monte Carlo-Markov Chains (MCMC) by sampling the posteriors. This package is based on the Affine-Invariant Ensemble Sampler \cite{GoodmanWeare}. To plot the results, we used the package getdist \cite{Getdist}, which uses the method of Kernel Density Estimation (KDE) in order to smooth the MCMC histograms.

\subsection{Results}
\begin{figure}[ht]
    \centering
    \includegraphics[width=.9\textwidth]{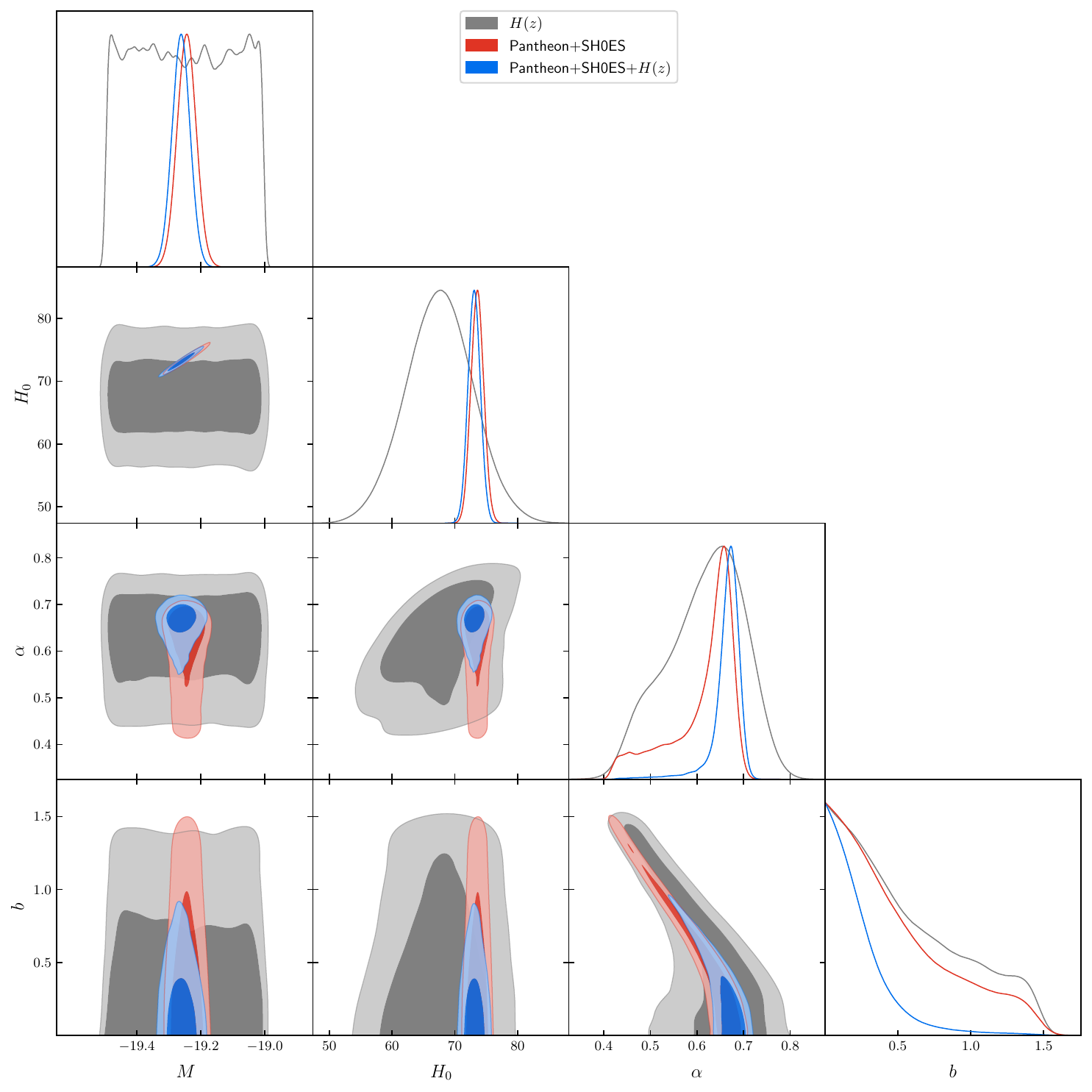}
    \caption{Constraints on $M$, $H_0$, $\alpha$, $b$, using the SNe Ia Pantheon+SH0ES data set, along with $H(z)$ from cosmic clocks.}
    \label{fig:triang-all}
\end{figure}

The constraints on the free parameters from all the datasets can be seen on Fig. \ref{fig:triang-all}.

\newpage

As can be seen on this Figure, the Pantheon+SH0ES dataset yields the strongest constraints. However, $H(z)$ data was essential in order to reduce the degeneracy between $\alpha$ and $b$ parameters. In fact, the degeneracy between $\alpha$ and $b$ was such that $b$ was superiorly limited only by the \textbf{flat} prior $b\in[0,1]$ both for Pantheon+SH0ES and $H(z)$ data. For the combination, however, it was superiorly limited at $b\sim1$ at 95\% c.l.

\begin{figure}[ht]
    \centering
    \includegraphics[width=.9\textwidth]{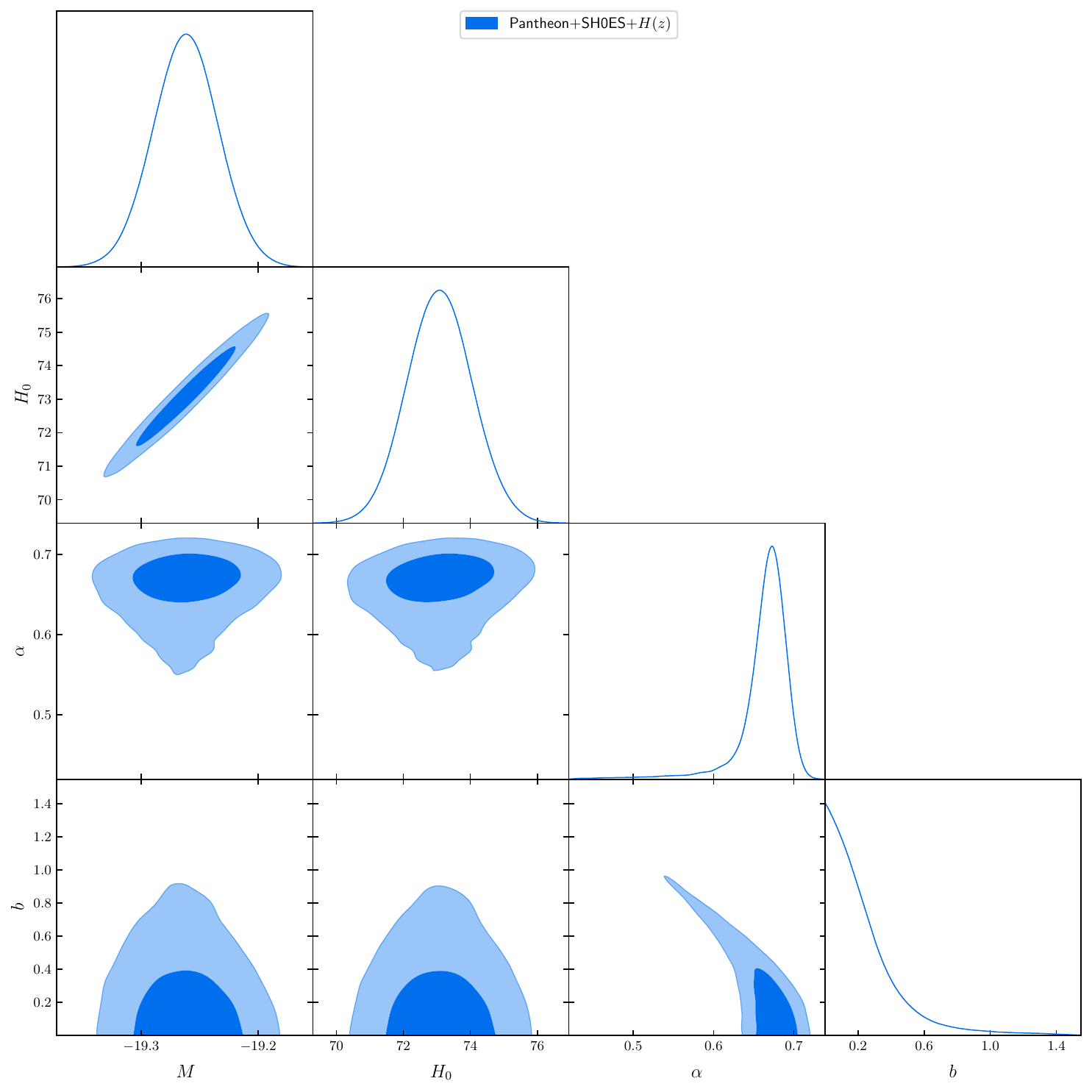}
    \caption{Combined Constraints on $\alpha$, $b$, $H_0$ and $M$, using the SNe Ia Pantheon+SH0ES+$H(z)$ from cosmic clocks.}
    \label{fig:triang-comb}
\end{figure}

In Fig. \ref{fig:triang-comb}, we see the combined results.

\newpage

As can be seen on this Figure, all parameters are well constrained by the joint analysis, except for $b$. Due to the degeneracy between $\alpha$ and $b$, the $b$ parameter is weakly constrained, although this degeneracy is reduced when combining with $H(z)$ data, as explained above. There is, also, a strong correlation between $H_0$ and $M$, the SNe Ia absolute magnitude, induced by Pantheon+SH0ES data. Nevertheless, this correlation also appears in analyses involving the standard cosmological model. The results for the free parameters at 1 and 2$\sigma$ c.l. can be seen at Table \ref{tab1}.

\begin{table}[ht] 
    \centering
   \begin{tabular} { l c c}
   
\hline
 Parameter & \ 68\% C.L & 95\% C.L.\\
\hline
{\boldmath$M              $} & $-19.262^{+0.028}_{-0.028}   $ & $-19.262^{+0.056}_{-0.056} $\\
{\boldmath$H_0            $} & $73.1^{+0.97}_{-0.97}        $ & $73.1^{+2.0}_{-1.9}        $\\
{\boldmath$\alpha         $} & $0.663^{+0.029}_{-0.012}     $ & $0.663^{+0.052}_{-0.068}   $\\
{\boldmath$b              $} & $< 0.266                     $ & $< 0.691                   $\\
\hline
\end{tabular}
    \caption{Combined constraints on the free parameters at 68\% and 95\% confidence levels (C.L.) from Pantheon+SH0ES+$H(z)$.}
    \label{tab1}
\end{table}

As can be seen from this Table, the $b$ parameter is well constrained at 1$\sigma$ c.l., if one considers that we have used only background data. It may indicate that the combination with clustering data or even other background data can provide strong constraints for this parameter.

%\begin{figure}
%\centering
%\includegraphics[scale=0.7]{SNeIa_Hz_b_alfa_star.pdf}
%\caption{Constraints on $\alpha$, $b$, $H_0$ and $M$, using the SNeIa Pantheon+SH0ES data set, along with $H(z)$ from cosmic clocks}
%\label{b_alpha_star}
%\end{figure}

%%%%%%%%%%%%%%%%%%%%%%%%%%%%%%%%%%%%%%%%%%%%%%%%%%%%%%%%%%%%%%%%%%%%%%%%%%%%%%%%%%%%%%%%%%%

\section{Final Comments and Conclusion}

In this paper, we have discussed based on Einstein's gravity theory, a new accelerating cosmology without vacuum, quintessence or extra dimensions. The late time accelerating phase is powered by the negative pressure (back reaction) due to the particle creation mechanism of massive, warm dark matter particles forming the reduced relativistic gas.   

In comparison with the standard cosmology ($\Lambda$CDM), the first advantage of our model is that the unknown dark sector (DE + DM) is reduced to only dark matter. In fact, he absence of vacuum energy density ($\Omega_{\Lambda}=0$) implies that the old cosmological constant problem and also the coincidence mystery  are absent in this framework. Secondly, it is known that the $\Lambda$CDM model is degenerated with the original creation cold dark matter (CCDM) cosmology \cite{LJO2010}, not only for background results, but also in the linear and nonlinear perturbative levels. It was shown  that the inclusion of the RRG warm dark matter broke the degeneracy between $\Lambda$CDM and CCDM cosmology, and, as such, a quasi-$\Lambda$CDM dynamics is obtained. In principle, this quasi-CCDM evolution cosmology may help to solve the $H_0$ and $S8$ tensions.   

On the other hand, the idea of relativistic dark matter particles as a significant component of the universe has problems nonetheless. It is widely believed that relativistic dark matter prevents large scale structure to be formed due to the free streaming process. However, as discussed in the introduction, some papers show that warm dark matter (WDM) can be consistent with structure formation while providing relativistic corrections to cosmology based on  the RRG model. In this context,  we were able to obtain an analytical model that recovers the pure RRG model when the number of particles is conserved $(\alpha=0)$ and the CCDM cosmology when only non-relativistic particles are considered ($b=0$). In other words, this new model $RRG + \alpha$ is a generalization of the CCDM(LJO) cosmology in the context of massive relativistic particles. 

%The dynamics of the universe in the $RRG + \alpha$ model slightly differs from the CCDM and $\Lambda$CDM models for small relativistic contributions. Consequently, these contributions might alleviate some problems of the standard model while being consistent with observations. 

It is also important to stress that the energy density of matter in the late phase of the universe is $\Omega_m \sim 1$ for models without dark energy. However, observations indicate that $\Omega_{m} \sim 0.3$. Nevertheless, the fraction of matter that clusters in galaxies is represented, in models with particle creation, by the effective energy density $\Omega_{meff}=\Omega_m - \alpha$ which is consistent with $\Omega_{meff}\approx 0.3$ as shown in CCDM model \cite{LJO2010}. Additionally, the remaining $0.7$ fraction of matter is said to be homogeneously distributed across the universe by the mechanism of particle creation in these cases. %Although the matter energy density in the $RRG + \alpha$ model is corrected by the warmness parameter $\Omega_m=1/\sqrt{1+b^2}$, its effective value $\Omega_{meff}=1/(\sqrt{1+b^2}) - \alpha$ must satisfy the observations.

In order to constrain the free parameters of the model, we have used background data as SNe Ia + Cepheids from Pantheon+SH0ES and $H(z)$ from Cosmic Chronometers. While the $b$ parameter is weakly constrained by this analysis, it enables distinctions from the standard $\Lambda$CDM model, differently of what happened with CCDM, where no distinction could be made, even at higher orders of density perturbations. Figures (\ref{fig:triang-all} and \ref{fig:triang-comb}) show that this combined data attribute the maximum probability for $b$ (for $b$ up to 1) at $b\ll1$, approximating the proposed model to CCDM and $\Lambda$CDM. This result is consistent with other background tests in models with conserved number of particles of RRG, as can be seen in \cite{Fabris2012}. On the other hand, these observations suggest that the transition from a decelerated to an accelerated regime of the universe happens slightly later in the proposed model at $z_t\approx 0.60$ (see Appendix), whereas in CCDM (LJO) model, $z_t\approx 0.71$. This result is expected, since the relativistic contributions increases the positive pressure from the gas, thereby intensifying the deceleration regime at high redshifts. Although $b$ also increases the negative pressure of particle creation, these contributions are less significant since they are stronger at high values of $z$, i.e. when $p_c$ is minimal. 

These results endorse the claim that this new model has mild differences in the cosmic dynamics, as compared to CCDM and $\Lambda$CDM. Therefore, these differences might clarify some of the problems plaguing the standard model. From the statistical analysis in section (\ref{analysis}), we can also verify the estimate of the effective energy density of matter according to the constraints on $\alpha$ and $b$. For instance, consider the upper limit at 1$\sigma$ c.l. constraint on $b$ at $b=0.266$ along with the best fit for $\alpha$ at $\alpha=0.663$ (see table \ref{tab1}) which yield $\Omega_{meff}\approx0.30$ as expected. Nevertheless, more data are required in order to better constrain the parameters within the $RRG + \alpha$ context. In fact, since cosmology is in the precision era, this model must also be confronted with observations in the perturbative level of cosmology. Based on some assumptions, the warmness parameter have already been constrained by analysing  density perturbations, however, the treatmente was applied for models with a conserved number of RRG particles. A perturbative extension of such analyses to the context of particle creation will be addressed in a forthcoming paper. 

\appendix

\section{The deceleration parameter \texorpdfstring{$q(z)$}{q(z)} and the transition redshift \texorpdfstring{$z_t$}{zt}}

\begin{figure}[ht]
    \centering
    \includegraphics[width=.6\textwidth]{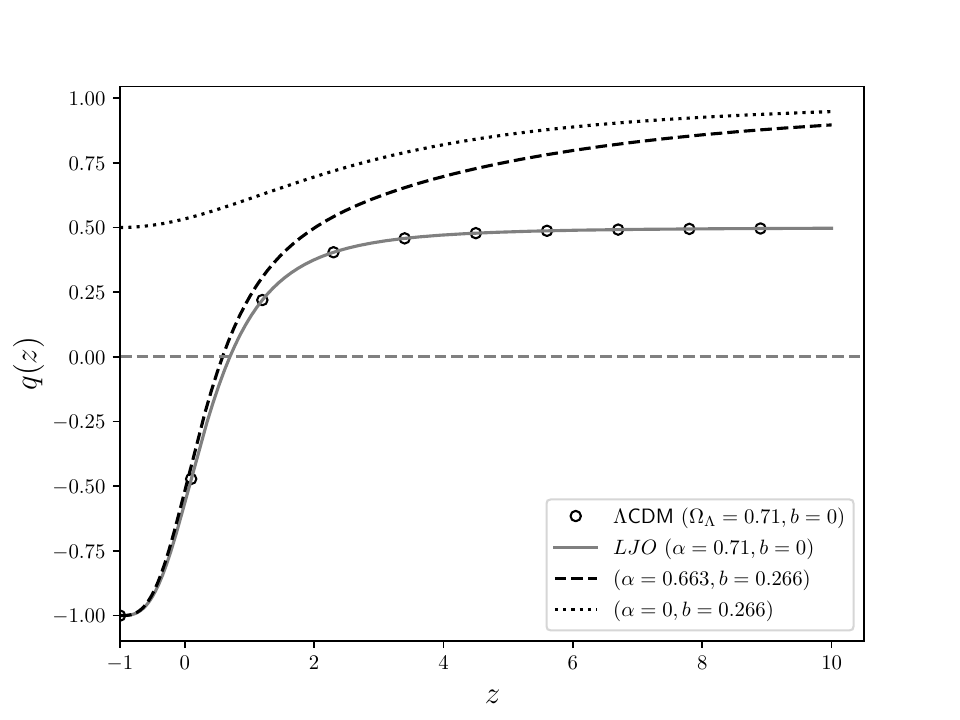}
    \caption{The deceleration parameter $q$ with respect to the redshift for different cosmological models. The solid grey line represents the CCDM (LJO) model, which expectedly coincides with the $\Lambda$CDM model represented by the black circles. The dashed line corresponds to the $q(z)$ of the RRG with particle creation for the values $\alpha=0.663$ and $b=0.266$ (see table \ref{tab1}). Finally, the dotted line indicates $q(z)$ of a model with RRG and absent of particle creation. }
    \label{figqz}
\end{figure}

Two useful dynamical physical parameters in cosmology are the deceleration parameter $q(z)$ and the value of the transition redshift, $z_t$. These parameters are very useful to discriminate cosmological models \cite{LJSM2012,PM2022,JesusEtAl19,JesusEtAl22}. In particular, $\Lambda$CDM and CCDM models predict the same values for both quantities. Since now we have two free parameter ($\alpha, b$), we show in this appendix that their values are slightly modified. The value of $q$ is defined by:
\begin{equation}
    q = - \frac{a\ddot{a}}{\dot{a}^2} = -\frac{\dot{H}}{H^2} -1.
\end{equation}
From the second Friedmann equation (\ref{EE2}), the Hubble parameter (\ref{Hzs}), and the expression above, the deceleration parameter can be obtained for the model proposed in this paper, and written in terms of the redshift, as
\begin{align}\label{qz}
    q\left(z\right)&=-1+2\left(1-\frac{\alpha}{\left(\frac{1}{\sqrt{1+b^{2}}}-\alpha\right)\left(1+z\right)^{3}+\alpha}\right)\times\nonumber\\
&\times\left(1-\frac{1}{4}\frac{1}{1+b^{2}\left(1+b^{2}\right)^{1/3}\left[\left(\frac{1}{\sqrt{1+b^{2}}}-\alpha\right)\left(1+z\right)^{3}+\alpha\right]^{2/3}}\right).
\end{align}
Note that, for $b=\alpha=0$, the deceleration parameter above yields $q=0.5$, which corresponds to a model with pure pressureless matter, as expected. On the other hand, for $\alpha\neq0$ and $b\neq0$ the deceleration parameter depends on $z$, and as $z\rightarrow\infty$ we notice that $q\rightarrow1$ which corresponds to the ultra-relativistic case. Finally, when $z\rightarrow-1$, i.e. $a\rightarrow\infty$, equation (\ref{qz}) approaches $q\approx-1$, which is the $q$ for vacuum. The dynamics of the universe for this model, regarding the deceleration parameter, can be visualized in figure (\ref{figqz}).

As in the $\Lambda$CDM and CCDM cosmologies, the model described above also transits from a decelerated to an accelerated expansion phase. This exchange occurs when $\ddot{a}=0$, or equivalently $q(z)=0$, at the transition redshift $z_{t}$. Hence, the transition redshift obtained by evaluating $q(z)=0$ in equation (\ref{qz}), is  
\begin{equation}\label{zt}
z_{t}=\left(\frac{2\alpha}{\frac{1}{\sqrt{1+b^{2}}}-\alpha}\right)^{1/3}\left\{ 1+\frac{b^{2}\left(1+b^{2}\right)^{1/3}\left[\left(\frac{1}{\sqrt{1+b^{2}}}-\alpha\right)\left(1+z_{t}\right)^{3}+\alpha\right]^{2/3}}{1+b^{2}\left(1+b^{2}\right)^{1/3}\left[\left(\frac{1}{\sqrt{1+b^{2}}}-\alpha\right)\left(1+z_{t}\right)^{3}+\alpha\right]^{2/3}}\right\} ^{-1/3}-1 .
\end{equation}
Notably, for $b=0$, the equation above recovers $z_t$ from the CCDM(LJO) model $z_t=[2\alpha/(1-\alpha)]^{1/3}  -1$ \cite{LJO2010}. However, for $b\neq0$, the transition redshift is modified by relativistic corrections. The value of $z_t$ is not easily obtained from the equation above due to higher order terms of $z_t$ in this expression. An approximation to the equation (\ref{zt}) can be obtained using a recurrence procedure, where at the zeroth-order, $z_t^{(0)}=z_t(b=0)$, and replacing this at the rhs of the equation above, we obtain the first order approximation:
\begin{equation}\label{ztapp}
z_{t}^{(1)}=\left(\frac{2\alpha}{\frac{1}{\sqrt{1+b^{2}}}-\alpha}\right)^{1/3}\left\{ 1+\frac{b^{2}\left(1+b^{2}\right)^{1/3}\left[\left(\frac{1}{\sqrt{1+b^{2}}}-\alpha\right)\left(\frac{2\alpha}{1-\alpha}\right)+\alpha\right]^{2/3}}{1+b^{2}\left(1+b^{2}\right)^{1/3}\left[\left(\frac{1}{\sqrt{1+b^{2}}}-\alpha\right)\left(\frac{2\alpha}{1-\alpha}\right)+\alpha\right]^{2/3}}\right\} ^{-1/3}-1 .
\end{equation}
The approximation above, for the values $\alpha=0.663$ and $b=0.266$ (see table \ref{tab1}), yields $z_{t}^{(1)}\approx 0.58$.

\begin{figure}[ht]
    \centering   
    \includegraphics[width=.4\textwidth]{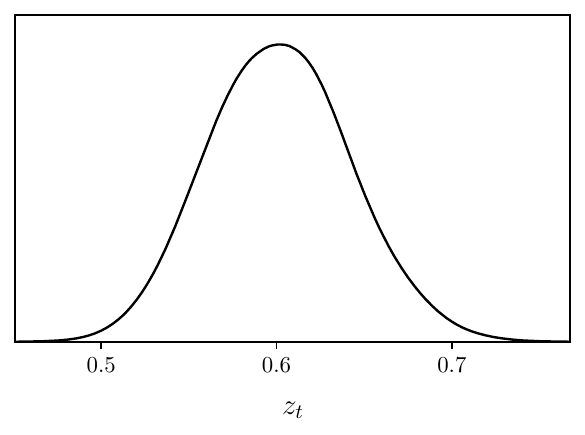}
    \caption{ The likelihood of the transition redshift $(z_t)$ for the RRG model in the context of particle creation given by the the combined observational data Pantheon+SH0ES \cite{Pantheon+} with $H(z)$ data from cosmic clocks \cite{MorescoEtAl20,MorescoEtAl22}, which constrains $z_t$ to $z_t=0.602^{+0.041 +0.085}_{-0.041 -0.079}$.}
    \label{figzt}
%\end{subfigure}   
\end{figure}

A more realistic estimation of $z_t$ can be obtained by calculating the likelihood of $z_t$ (see Figure \ref{figzt}) with the constraints on $\alpha$ and $b$ previously sampled from the combined Pantheon+SH0ES+$H(z)$ data, discussed in the previous section, and the theoretical prediction of $z_t$ in equation (\ref{zt}). These results yield $z_t=0.602^{+0.041}_{-0.041}$ for a 68\% C.L. and $z_t=0.602^{+0.085}_{-0.079}$ for a 95\% C.L.. As expected, this transition redshift is slightly smaller than $z_t(LJO)\approx 0.71$ in the CCDM(LJO) model \cite{LJO2010}, since the relativistic contribution of the gas increases the positive pressure (\ref{EoSRRG}). In other words, the pressure, which is supposed to be zero for massive particles, receives a positive contribution from the relativistic particles that slows the expansion of the universe, compared to a standard matter-dominated era, and retards the transition for an accelerated regime. It is interesting to notice that $b$ also contributes to the negative creation pressure (see \ref{Pc1} and \ref{EoSRRG}), although this contribution is less relevant due to the fact that $\Gamma$ is only significant when $\rho_{m}$ becomes small (see equation \ref{gammadm}).

\begin{acknowledgments}
PWRL is financially supported by the National Council for Scientific and Technological Development (CNPq) under grant 140100/2021-0. JFJ is partially supported by CNPq under grant 314028/2023-4. JASL is also partially supported by the Brazilian agencies, CNPq under Grant 310038/2019-7, CAPES
(grant 88881.068485/2014) and FAPESP (LLAMA Project 11/51676-9). 
\end{acknowledgments}

%%%%%%%%%%%%%%%%%%%%%%%%%%%%%%%%%%%%%%%%%%%%%%%%%%%%%%%%%%%%%%%%%%%%%%%%%%%%%%%%%%%%%%%%%%%%%%%%%%%%


\bigskip

\begin{thebibliography}{99}

\bibitem{supernovae} M.~Hicken et~al.
Astrophys. J. {\bf 700}, 1097 (2009); R. Amanullah {\it et~al.}, Astrophys. J. {\bf 716}, 712 (2010).

\bibitem{Suzuki:2011hu} N.~Suzuki, D.~Rubin, C.~Lidman, G.~Aldering, R.~Amanullah,
K.~Barbary, L.~F.~Barrientos and J.~Botyanszki {\it et al.},
Astrophys.\ J.\  {\bf 746}, 85 (2012).

\bibitem{Pantheon+} D.~Brout, D.~Scolnic, B.~Popovic, A.~G.~Riess, J.~Zuntz, R.~Kessler, A.~Carr, T.~M.~Davis, S.~Hinton and D.~Jones, \textit{et al.} Astrophys. J. \textbf{938}, no.2, 110 (2022).
[arXiv:2202.04077 [astro-ph.CO]].

\bibitem{AL199} L. Krauss, Astrophys. J. 480, 466 (1997), J. S. Alcaniz and J. A. S. Lima, Astrophys. J. 521 L87 (1999); J. A. S. Lima and J. S. Alcaniz, MNRAS 317, 893, (2000); 

\bibitem{CS2004} J. V. Cunha and R. C. Santos, IJMPD \textbf{13}, (07) 1321 (2004); J. S. Alcaniz, J. A. S. Lima and J. V. Cunha, MNRAS 340, L39 (2003); A. C. S. Friaça, J. S. Alcaniz  and J. A. S. Lima, MNRAS 362 (4), 1295 (2005), 


\bibitem{Cluster1} S. W. Allen, S. Ettori, and A. C. Fabian, Mon. Not. R. Astron. Soc. {\bf 324}, 877 (2002);  J. A. S. Lima, J.  V. Cunha and J. S. Alcaniz, Phys. Rev. D {\bf 68}, 023510 (2003), astro-ph/0303388; S. W. Allen et al., Mon. Not. R. Astron. Soc. {\bf 383}, 879 (2008). 

\bibitem{Cluster2} R. Jimenez, L. Verde, T. Treu and D. Stern, Astrophys. J. \textbf{593}, 622 (2003); S. Basilakos and J. A. S. Lima,  Phys. Rev. D 82 (2), 023504 (2010); S. Basilakos, M. Plionis and J. A. S. Lima Rev. D 82 (8), 083517 (2010).

\bibitem{SDSS2005} D. J. Eisenstein \textit{et al.}, SDSS Collaboration, Astrophys. J. \textbf{633}, 560 (2005); A. G. Sanchez \textit{et al.}, MNRAS \textbf{425}, 415 (2012). 

\bibitem{DESI2024} A. Heinensen, Y.-Z. Li and D. L. Wiltshire, JCAP 03,003 (2019); A. G. Adame, J. Aguilar, S. Ahlen, S. Alam \textit{et al.}, arXiv:2404.03002v3 [astro-ph.CO].

\bibitem{Planck2019} T. M. C. Abbott, et al., Astrophys. J. Lett. 872, L30 (2019).

\bibitem{Planck2020} Aghanim, et al., Planck results 2018, Astron. Astrophys. 641, A6 (2020).

\bibitem{JimenezLoeb02}
R.~Jimenez and A.~Loeb, Astrophys. J. \textbf{573}, 37 (2002),
[arXiv:astro-ph/0106145 [astro-ph]].

\bibitem{MorescoEtAl20} M.~Moresco, R.~Jimenez, L.~Verde, A.~Cimatti and L.~Pozzetti, Astrophys. J. \textbf{898}, no.1, 82 (2020). [arXiv:2003.07362 [astro-ph.GA]].

\bibitem{MorescoEtAl22} M.~Moresco, L.~Amati, L.~Amendola, S.~Birrer, J.~P.~Blakeslee, M.~Cantiello, A.~Cimatti, J.~Darling, M.~Della Valle and M.~Fishbach, \textit{et al.}, Living Rev. Rel. \textbf{25}, 6 (2022). 

\bibitem{review} P. J. E. Peebles and B. Ratra  Rev. Mod. Phys. {\bf 75}, 559 (2003); T. Padmanabhan,  Phys. Rept. {\bf 380}, 235 (2003);  J. A. S. Lima, Braz. J. Phys. {\bf 36}, 1109 (2004);  E. J. Copeland, M. Sami and   S. Tsujikawa, Int. J. Mod. Phys. D {\bf 15}, 1753 (2006); J.~A.~Frieman, M. S. Turner and D. Huterer, Ann. Rev. Astron. Astrophys. {\bf 46}, 385 (2008); M. Bartelmann, Rev. Mod. Phys. {\bf 82}, 331 (2010).

\bibitem{SW1989}  Ya. B. Zeldovich, JETP Lett. 6, 316 (1967); A. Zee, in High Energy Physics, Proceedings of the 20th Annual Orbis Scientiae, edited by B. Kursunoglu, S. L. Mintz, and A. Perlmutter, (Plenum, New York, 1985); S. Weinberg, Rev. Mod. Phys. 61, 1 (1989).
\bibitem{PC03}  P. J. Steinhardt, in: V. L. Fitch, D. R. Marlow, M. A. E. Dementi (Eds.), Critical Problems in Physics, Princeton University, Princeton, NJ, 1997; P. J. Steinhardt, Philos. Trans. R. Soc. Lond. A 361,  2497 (2003).

\bibitem{Riess2019} A. G. Riess, \textit{et al.}, Astrophys. J. \textbf{876},  85 (2019).

\bibitem{B2019} S. Birrer, \textit{et al.}, Mon. Not. R. Astron. Soc. \textbf{484}, 4726 (2019).

\bibitem{LV07} J. A. S. Lima, J. V. Cunha, Astrophys. J. Lett. \textbf{781},  L38 (2014);  J. V. Cunha, L. Marassi and J. A. S. Lima, Mon. Not. R. Astron. Soc. \textbf{379}, L1 (2007).

\bibitem{kids} KiDS Collaboration, M. Asgari, \textit{et al.},  Astron. Astrophys. \textbf{645}, A104 (2021).

\bibitem{DS2017} J. A. S. Lima and D. Singleton, Eur. Phys. J. C \textbf{77}, 855 (2017). 

\bibitem{LH} M. Ozer and M. O. Taha, Phys. Lett. B {\bf 171}363  (1986);
W. Chen,Y.-S. Wu, Phys. Rev. D {\bf 41}, 695 (1990); J. C. Carvalho, J. A. S. Lima and I. Waga, Phys. Rev. D {\bf 46}, 2404 (1992); I ~Waga, Astrophys. J. {\bf 414}, 436 (1993); J. A. S. Lima and J. M. F. Maia, Phys. Rev.
D \textbf{49}, 5597 (1994); J. A. S. Lima,  Phys. Rev. D  \textbf{54} (4), 2571 (1996); P. E. M. Almeida, R. C. Santos and J. A. S. Lima, Universe \textbf{10} (9), 362 (2024).

\bibitem{Inter} E. Abdalla \textit{et al.}, JHEP 34, 49 (2022); A. A. Costa, X. D. Xu, B. Wang, E. G. M. Ferreira, E. Abdalla Phy. Rev. D \textbf {89} (10), 103531 (2014); J. A. S. Lima, A. I. Silva and S. M. Viegas, MNRAS \textbf{312}, 747 (2000). 

\bibitem{SF2010} T. P. Sotiriou and V. Faraoni, Rev. Mod. Phys. 82, 451 (2010); S. Capozziello and M. De Laurentis, Phys. Rep. 509, 167 (2011).

\bibitem{LSS2008}  J. A. S. Lima, F. E. Silva and R. C. Santos, Classical  Quantum Gravity {\bf 25}, 205006 (2008), arXiv:0807.3379;  G. Steigman, R. C. Santos and J. A. S. Lima, J. Cosmol. Astropart. Phys. {\bf 06} 033 (2009), arXiv:0812.3912.

\bibitem{LBC2012} J. P. Mimoso and D. Pavón, Phys. Rev. D 87, 047302 (2013), arXiv:1302.1972;  
 J. A. S. Lima, S. Basilakos and F. E. M. Costa, Phys. Rev. D {\bf 86}, 103534 (2012), arXiv:1205.0868; S. Basilakos and J. A. S. Lima, Phys. Rev. D {\bf 82}, 023504 (2010), arXiv:1003.5754

\bibitem{LJO2010} J. A. S. Lima, J. F. Jesus and F. A. Oliveira, J. Cosmol. Astropart. Phys. {\bf 11}  027 (2010), arXiv:0911.5727 

\bibitem{Pert1} R. O. Ramos, M. V. dos Santos and I. Waga, Phys. Rev. D {\bf 89}, 083524 (2014); M. V. Santos, I. Waga, R. O. Ramos, Phys. Rev. D {\bf 90}, 127301 (2014).

\bibitem{Pert2} J. A. S. Lima, R. C. Santos, and J. V. Cunha,  JCAP, (03), 027 (2016).

\bibitem{TL2023} S. R. G  Trevisani and J. A. S. Lima,  EPJC \textbf{83} (3), 1-17 (2023);  J. A. S. Lima, S. R. G. Trevisani and R. C. Santos, Phys. Lett. B  \textbf{820}, 136575 (2021). 

\bibitem{EKO2024} E. Elizalde, M. Khurshudyan,  S. D. Odintsov,  EJPC \textbf{84}, 782 (2024); V. H. Cárdenas, M. Cruz and S. Lepe, PDU \textbf{37}, 101122 (2022).

\bibitem{Hipolito2018} W. S. Hipólito-Ricaldi, \textit{et al.}, EPJC, \textbf{78}, 365 (2018).

\bibitem{Pordeus2019} L. G. Medeiros, L. Gouvêa, G. Pordeus-da-Silva and R. C. Batista, JCAP \textbf{06}, 043 (2019).

\bibitem{Pordeus2021} G. Pordeus-da-Silva, R. C. Batist, and L. G. Medeiros, JCAP \textbf{11}, 062 (2021) .

\bibitem{PRI89} I. Prigogine {\it et al.}, Gen. Rel. Grav. {\bf 21}, 767 (1989).

\bibitem{CLW92} M. O. Calvao, J. A. S. Lima and I. Waga, Phys. Lett. A {\bf 162}, 223 (1992). See also, J. A. S. Lima, M. O. Calv\~{a}o and I. Waga, ``Cosmology, Thermodynamics and Matter Creation'', {\it Frontier Physics, Essays in
Honor of Jayme Tiomno}, World Scientific, Singapore (1990), also in arXiv:0708.3397

\bibitem{LG92} J. A. S. Lima and A. S. M Germano, Phys. Lett. A {\bf 170}, 373 (1992)
\bibitem{zeld70} Ya. B. Zeldovich, JETP Lett. {\bf 12} 307 (1970).

\bibitem{PZ93} I. Waga,  R. C. Falcao and R. Chanda, Physical Review D [\textbf{33}, 1839 (1987);  J. A. S. Lima, R. Portugal and I. Waga, Phys. Rev. D {\bf 37}
 2755 (1988).

 \bibitem{PT} N.D. Birrell and P.C.W. Davies, \textit{Quantum Fields in Curved Space}, Cambridge University Press, Cambridge U.K. (1989); L. Parker and D. Toms, \textit{Quantum Field Theory in Curved Spacetimes}, Cambridge University Press, Cambridge U.K. (2009).

\bibitem{PDU1} B. Hoeneisen, Phys. Dark Universe \textbf{46},  101643 (2024).

\bibitem{PDU2} C. Yi Tan, A. Dekker, A. Drlica-Wagner, arXiv:2409.18917 [astro-ph.CO] (2024).

\bibitem{Horiuchi2016} S. Horiuchi, \textit{et al.}  MNRAS \textbf{456},  4346 (2016).

\bibitem{Jutner1911} J. L. Synge, The Relativistic Gas (North-Holland, Amsterdam, 1957); S. R. de Groot and P. Mazur, Relativistic Kinetic Theory, North Holland (1980), Amsterdam (1980); W. Pauli, \textit{Theory of Relativity}, Dover Books in Physics (1981).

\bibitem{sakharov1966} A. D. Sakharov, Sov. Phys. JETP 22: 241 (1966).

\bibitem{Peixoto2005} G. De Berredo-Peixoto, I. L. Shapiro and F. Sobreira,  
MPLA, \textbf{20}, 2723 (2005).

\bibitem{Fabris2009} J. C. Fabris,  I. L. Shapiro, and F. Sobreira,  
JCAP \textbf{02}, 001 (2009).

\bibitem{Fabris2012} C. Fabris, I. L. Shapiro, and A. M. Velasquez-Toribio, Phys. Rev. D {\textbf{85}}, 02350 (2012); L. G. Medeiros, Modern Physics Letters A 27, 33, 125019 (2012).
%\textit{Cosmological analytic solutions with reduced relativistic gas}. 

\bibitem{Slepian2018} Z. Slepian, S. K. N. Portillo,  MNRAS \textbf{478}, 516 (2018).

\bibitem{Ruiz2020} J. A. Agudelo Ruiz, J. C. Fabris, A. M. Velasquez-Toribio, and I. L. Shapiro,  Gravitation and Cosmology  \textbf{26} , 316 (2020). 

\bibitem{Reis2018} S. C. dos Reis, I. L. Shapiro, EPJC,  \textbf{78}, 145 (2018). 

\bibitem{Rapetti2008} D. Rapetti, S. W. Allen, and A. Mantz, MNRAS  \textbf{388}, 1265 (2008).

\bibitem{Rod2019} R. F. L. Holanda, R. S. Gonçalves, J. E. Gonzalez and J. S. Alcaniz, JCAP, \textbf{11} 032 (2019).

\bibitem{Q2024} L. Quiu \textit{et al.} A\&A, 687, A1 (2024).

\bibitem{ForemanMackey13} Foreman-Mackey, D. W. Hogg, D. Lang and J. Goodman, Publications of the ASP
\textbf{125},  306 (2013), [arXiv:1202.3665 [astro-ph.IM]].

\bibitem{GoodmanWeare} J. Goodman and J. Weare, Comm. Appl. Math. Comput. Science \textbf{5}, 33, 65 (2010).

\bibitem{Getdist} A. Lewis [arXiv:1910.13970 [astro-ph.IM]].

\bibitem{LJSM2012} J. A. S. Lima, J. F. Jesus, R. C. Santos, M. S. S. Gill, arXiv:1205.4688 (2012); J. V. Cunha  and J. A. S. Lima, MNRAS, \textbf{390}, 210 (2008).  

\bibitem{PM2022} P. Mukherjee and N. Banerjee, PDU \textbf{36}, 100998 (2022).

%\cite{Jesus:2019nnk}
\bibitem{JesusEtAl19}
J.~F.~Jesus, R.~Valentim, A.~A.~Escobal and S.~H.~Pereira,
%``Gaussian Process Estimation of Transition Redshift,''
JCAP \textbf{04} (2020), 053
%doi:10.1088/1475-7516/2020/04/053
[arXiv:1909.00090 [astro-ph.CO]].
%64 citations counted in INSPIRE as of 17 Feb 2025

%\cite{Jesus:2022xwb}
\bibitem{JesusEtAl22}
J.~F.~Jesus, D.~Benndorf, A.~A.~Escobal and S.~H.~Pereira,
%``From Hubble to snap parameters: a Gaussian process reconstruction,''
Mon. Not. Roy. Astron. Soc. \textbf{528} (2024) no.2, 1573-1581
%doi:10.1093/mnras/stae120
[arXiv:2212.12346 [astro-ph.CO]].
%6 citations counted in INSPIRE as of 17 Feb 2025

\end{thebibliography}
\end{document}